\documentclass[aps,showpacs,amsmath,amssymb,twocolumn,prb,longbibliography,superscriptaddress,notitlepage,floatfix,10pt]{revtex4-2}

\usepackage[dvips]{graphicx}
\usepackage{bm, bbm,ascmac,amsmath,amssymb,amsthm,mathrsfs,amsfonts,dsfont}
\usepackage{epsfig}
\usepackage{braket}
\usepackage{enumerate}
\usepackage{xcolor}
\usepackage{float}
\usepackage{algorithm}
\usepackage{algorithmic}
\usepackage{comment}
\usepackage{appendix}
\usepackage{here}
\usepackage{tabularx}
\usepackage{dcolumn}
\usepackage[caption=false]{subfig}
\usepackage{adjustbox}
\usepackage[hidelinks]{hyperref}
\usepackage{bookmark}
\usepackage{makecell}

\newtheorem{theorem}{Theorem} 

\usepackage{scalerel}

\iftrue
  \usepackage{marginnote}
\fi
\begin{document}

\title{Spatial-Order Hierarchy of Time-Dependent Exchange-Correlation Potential}
\author{Naoki Negishi}
\email{negishi@roma2.infn.it}
\affiliation{Dipartimento di Fisica, Università di Roma Tor Vergata, Via della Ricerca Scientifica 1, 00133 Rome, Italy}
\affiliation{INFN, Sezione di Roma Tor Vergata, Via della Ricerca Scientifica 1, 00133 Rome, Italy}

\begin{abstract}
Exact time-dependent density-functional theory (DFT) separates the exchange-correlation potential into interaction and kinetic-correlation components, but the structural relation between them remains unclear. We establish a representability constraint based on the off-diagonal expansion of the one-electron reduced density matrix (1RDM) relative to a chosen reference dynamics. A non-zero linear off-diagonal term of the 1RDM generates a current density beyond the representable space of the reference, while a nontrivial quadratic off-diagonal structure activates the kinetic-correlation component. The resulting hierarchy provides a criterion for assessing, from the spatial locality of electronic correlation, whether a given exchange-correlation functional or a more general electron-correlation approximation remains adequate. In a one-dimensional two-electron correlation quench, retaining the exact interaction component while setting the kinetic component to zero suppresses the density broadening of the exact evolution. These results establish a spatial-order hierarchy among the force components governing the density dynamics and provide an exact constraint for the construction and assessment of nonadiabatic approximations.
\end{abstract}
\maketitle

\section{introduction}
Time-dependent density functional theory (TDDFT)\cite{HKthm,KSthm,parr1989density,burke2012perspective,runge1984density,TKatoiKS,RvLPRA1994,PRLGrossKohn1985} provides, in principle, the exact time-dependent electron density, thereby reducing the description of electron dynamics to the electron-density space. Owing to its computational advantages over wavefunction-based methods, TDDFT has a broad range of applications, from optical processes involving electronic states~\cite{CASIDA1996391,PhysRevLett_Gross_1996,BAUERNSCHMITT1996454} to high-order harmonic generation and ionization~\cite{RvLPRA1994,PhysRevLett_NSDI_2005,PhysRevA_TDRNOT_2014,GrossPRL2010,GrossPRL2017,PhysRevB_Yabana}. On the other hand, in most practical applications, the exchange-correlation functional must be approximated, and the limitations of these approximations have been demonstrated in both equilibrium and nonequilibrium settings. In particular, nonadiabatic effects can significantly affect the accuracy of simulations of many-electron dynamics induced by free-electron injection~\cite{Kvaal2010} or strong-field excitation~\cite{RvLPRA1994,KATO2004533,TSato_2014}.

In TDDFT, the exchange-correlation potential representing many-electron effects must account for memory effects in the density dynamics, but constructing the corresponding memory kernel as a functional of the density remains difficult~\cite{KinC-XC-JCP2014}. Most applications therefore employ the adiabatic approximation for the exchange-correlation potential, evaluating the ground-state exchange-correlation functional at the instantaneous time-dependent density. Consequently, how these functionals reflect physical constraints on nonadiabatic dynamics remains less well understood than the corresponding constraints in low-order optical response.

Recent work has sought exact structural decompositions of the time-dependent exchange-correlation potential~\cite{XC-Scatter-PRL2017}. Within a hydrodynamic formulation based on the electron density, the exchange-correlation potential is partitioned into a kinetic-correlation term and an electron-interaction exchange-correlation term. Luo et al. numerically demonstrated the distinct dynamical roles of these terms~\cite{KinC-XC-JCP2014}. However, the mathematical structure and physical constraints underlying this separation have not been established analytically, so its interpretation has remained primarily numerical. Furthermore, the formulation of these two components is restricted to one-dimensional systems. In a general three-dimensional system, the transverse component of the physical current density does not necessarily coincide with the transverse component of the current density generated by the Kohn--Sham system~\cite{KSthm}. Consequently, the driving force associated with the transverse current component, which is not encoded in the density-based exchange-correlation functional, must be specified separately. The resulting force decomposition therefore does not constitute a strict decomposition of the terms making up the scalar exchange-correlation potential.

In our work, we revisit how the decomposition into the interaction and kinetic components is defined for realistic three-dimensional systems. We instead introduce force fields for the two components, which include not only the gradient of the scalar potential but also the rotational contribution associated with the vector potential. After establishing that the decomposition is well defined, we identify a constraint linking the kinetic-correlation term and the electron-interaction exchange-correlation term, based on the off-diagonal structure of the one-electron reduced density matrix (1RDM)~\cite{DMFT1,DMFT2,RDM}, whose dependence on the relative coordinate reflects quantum nonlocal correlations. By expanding the off-diagonal part of the 1RDM to first order in the relative coordinate, we show that many-body effects enter nontrivially only through the electron-interaction exchange-correlation term. This mathematical constraint shows that the kinetic-correlation contribution first appears at second order in the relative-coordinate expansion of the 1RDM, which is the lowest order at which spatial quantum nonlocal correlations contribute. In a one-dimensional two-electron model, we compare the equations of motion with and without the kinetic-correlation term, and then show that this term is essential for describing the nonstationary evolution of the electron density.

  \section{Revisit classification of the exchange-correlation potential}
We first review the decomposition of exchange-correlation potential introduced by Luo et. al.~\cite{KinC-XC-JCP2014} Let the time-dependent anti-symmetric wavefunction in an $N$-electron system be $\Psi({\bf r}_1\sigma_1,{\bf r}_2\sigma_2,\cdots,{\bf r}_N\sigma_N,t)$. Here, variables ${\bf r}_i$ and $\sigma_i$ denote the spatial and spin coordinates of the $i$-th electron, respectively, and $t$ denotes time.

It is known that the second time derivative of the electron density $n({\bf r},t)=N\int d\sigma_1\cdots\int d\sigma_N$ $\int d{\bf r}_2\cdots d{\bf r}_N$ $ |\Psi({\bf r}\sigma_1,{\bf r}_2\sigma_2,\cdots,{\bf r}_N\sigma_N,t)|^2$ can be written for the interacting system as follows~\cite{KinC-XC-JCP2014}:
\begin{equation}
\begin{split}
 \frac{\partial^2}{\partial t^2}n({\bf r},t)=\nabla\cdot\bigr(n\nabla v_{\rm ext}\bigl)+i\nabla\cdot\braket{\Psi(t)|[\hat{\bf j}({\bf r}),\hat{T}+\hat{W}]|\Psi(t)},   
\end{split}
~\label{eq:densityEOM}
\end{equation}
using atomic units $(\hbar=c=e=m_{\rm e}=1)$. The real-valued function $v_{\rm ext}({\bf r}_i,t)$ denotes one-electron external potential. The operators $\hat{T}=\sum_{i=1}^{N}-\frac{1}{2}\nabla_i^2$ and $\hat{W}=\sum_{i>j} w_{\rm ee}(|{\bf r}_i-{\bf r}_j|)$ represent the total kinetic-energy term and the two-electron interaction-potential term, respectively. The operator $\hat{\bf j}({\bf r})$ is the current-density operator, defined by $\hat{\bf j}({\bf r})=\frac{1}{2i}\sum_{i=1}^N\left(\delta({\bf r}-{\bf r}_{i})\nabla_i-({\rm h.c.})\right)$.

Based on time-dependent Kohn-Sham (TDKS) theory in Ref.~\cite{KinC-XC-JCP2014}, we obtain the equivalent Kohn–Sham expression of the second-time derivative of the electron density as follows:
\begin{equation}
    \frac{\partial^2}{\partial t^2}n({\bf r},t)=\nabla\cdot\bigr(n\nabla (v_{\rm ext}+v_{\rm H}+v_{\rm xc}+v_{s}^{T})\bigl).~\label{eq:densityEOMKS}
\end{equation}
The constituent $v_{\rm H}$ is the Hartree potential defined by
\begin{equation}
    v_{\rm H}({\bf r},t)=\int d{\bf r'} w_{\rm ee}(|{\bf r}-{\bf r'}|)n({\bf r'},t),
\end{equation}
and $v_{\rm xc}$ denotes the exchange-correlation potential in KS theory~\cite{KSthm}. The remaining term $n\nabla v_{s}^{T}$ stems from the kinetic-energy density in the KS system and is defined as 
\begin{align}
    \nabla \cdot(n({\bf r},t)\nabla v_{s}^{T}({\bf r},t))
    &=-\frac{1}{4}(\nabla^2-\nabla'^2)^2\gamma_{s}({\bf r},{\bf r'},t)|_{{\bf r}={\bf r'}}.\label{eq:vst}
\end{align}
Here, the 1RDM of the KS system is defined as $\gamma_{s}({\bf r},{\bf r'},t)=\int d\sigma\sum_{k=1}^m\phi_k({\bf r}\sigma,t)\phi_k^*({\bf r'}\sigma,t)$ by using TDKS spin-orbitals $\{\phi_{k}\}$. 

Using the equivalence of Eqs.~\eqref{eq:densityEOM} and~\eqref{eq:densityEOMKS}, the exchange-correlation potential $v_{\rm xc}$ is separated, with explicit dependence on $w_{\rm ee}$, into the term $v_{\rm xc}^{W}$ and the remaining term $v_{\rm c}^{T}$ in the previous study~\cite{KinC-XC-JCP2014}. This decomposition, however, is limited in one-dimensional modeling. In the present study, we need to care for the rotational fields in the extention to three-dimensional model from the previous one-dimensional definition of $v_{\rm xc}^{W}$ and $v_{\rm c}^{T}$.

\section{Force fields of the interaction and kinetic-correlation components}

We therefore revisit the definition of the two decomposed components of the exchange-correlation potential. We first define the force fields associated with the interaction and kinetic components, ${\bf F}_{\rm xc}^{W}({\bf r},t)$ and ${\bf F}_{\rm c}^{T}({\bf r},t)$, respectively, as follows:
\begin{equation}
\begin{split}
{\bf F}_{\rm xc}^{W}({\bf r},t)=-\int d{\bf r'}n({\bf r'},t)g_{\rm xc}({\bf r},{\bf r'},t)\nabla w_{\rm ee}(|{\bf r}-{\bf r'}|),  \label{eq:wxc}
\end{split}
\end{equation}
and
\begin{equation}
\begin{split}
    &{\bf F}_{\rm c}^{T}({\bf r},t)\\
    &=\frac{1}{4n({\bf r},t)}(\nabla-\nabla')(\nabla^2-\nabla'^2)\bigl(\gamma({\bf r},{\bf r'},t)-\gamma_{s}({\bf r},{\bf r'},t)\bigr)|_{{\bf r}={\bf r'}}\\
&={\bf F}_{\rm c}^{T}[\hat{\gamma}(t),{\bf r}],\label{eq:vcT}
    \end{split}
\end{equation}
where $g_{\rm xc}({\bf r},{\bf r'},t)$ and $\gamma({\bf r},{\bf r'},t)$ are the pair-correlation function and the 1RDM of the interacting system, respectively, defined as follows:
\begin{equation}
\begin{split}
    1+g_{\rm xc}({\bf r},{\bf r'},t)
    &=\frac{N(N-1)}{n({\bf r},t)n({\bf r'},t)}
    \int d\sigma_1\cdots d\sigma_N\\
    &\times\int d{\bf r}_3\cdots d{\bf r}_N|\Psi({\bf r}\sigma_1,{\bf r'}\sigma_2,
    \cdots,{\bf r}_N\sigma_N,t)|^2,
\end{split}
\end{equation}
and
\begin{align}
\gamma({\bf r},{\bf r'},t)
&=N\int d\sigma_1\cdots d\sigma_N
\int d{\bf r}_2\cdots d{\bf r}_N\notag\\
&\quad\times\Psi({\bf r}\sigma_1,{\bf r}_2\sigma_2,\cdots,{\bf r}_N\sigma_N,t)\notag\\
&\quad\times\Psi^*({\bf r'}\sigma_1,{\bf r}_2\sigma_2,\cdots,{\bf r}_N\sigma_N,t).
\end{align} 

Reference~\cite{KinC-XC-JCP2014} describes how the two exchange-correlation potential components can be uniquely determined by spatially integrating the corresponding force fields in one dimension. In three dimensions, however, the two force fields arising from different sources need not be curl-free; indeed, $\nabla\times{\bf F}_{\rm xc}^{W}\not=0$ and $\nabla\times{\bf F}_{\rm c}^{T}\not=0$. Representing these force fields in terms of potentials therefore requires introducing vector potentials for the interaction and kinetic components. In the present work, however, we do not need to introduce these potentials to derive the constraint between the two components discussed in the next section.

We next consider the boundary conditions for the decomposed force fields. Because the total exchange-correlation force field is uniquely determined once appropriate boundary conditions are imposed, specifying the force field ${\bf F}_{\rm xc}^W$ uniquely determines ${\bf F}_{\rm c}^T$. For a finite system with open boundary conditions, the interaction force field vanishes at infinity, so that $\lim_{|{\bf r}|\rightarrow\infty}{\bf F}_{\rm xc}^W=0$. For periodic boundary conditions with the unit lattice vector ${\bf L}$, the periodicity condition ${\bf F}_{\rm xc}^W({\bf r}+{\bf L},t)={\bf F}_{\rm xc}^W({\bf r},t)$ ensures the uniqueness of ${\bf F}_{\rm c}^T$.

When decomposing the exchange-correlation contribution, the decomposed force fields must transform consistently under gauge transformations of the interacting wavefunction and the Kohn--Sham reference system. The $N$-electron probability density is given by the squared modulus of the wavefunction, $|\Psi|^2$. Thus, quantities defined in terms of $|\Psi|^2$, such as the electron density and pair-correlation function, are gauge invariant. Therefore, the interaction component ${\bf F}_{\rm xc}^W$ is gauge invariant. The 1RDM, on the other hand, transforms under a gauge transformation characterized by $\chi({\bf r},t)$ as $\gamma({\bf r},{\bf r'},t)\rightarrow e^{i\chi({\bf r},t)-i\chi({\bf r'},t)}\gamma({\bf r},{\bf r'},t)$; the KS 1RDM transforms in the same way: $\gamma_s({\bf r},{\bf r'},t)\rightarrow e^{i\chi({\bf r},t)-i\chi({\bf r'},t)}\gamma_s({\bf r},{\bf r'},t)$. Nevertheless, under a general local gauge transformation, the kinetic operator must be transformed simultaneously with the interacting and Kohn--Sham 1RDMs. The corresponding gauge-covariant form of the kinetic-correlation force therefore represents the same physical force field. Equation~\eqref{eq:vcT} is its representation in the gauge with a vanishing vector potential.

\section{Spatial-order hierarchy of the exchange-correlation force components}

We establish an explicit relation between the spatial order of the off-diagonal elements of the 1RDM and the two components of the decomposition introduced above. The electron-interaction exchange-correlation force field ${\bf F}_{\rm xc}^{W}$ and the kinetic-correlation force field ${\bf F}_{\rm c}^{T}$ need not be independent functionals and generally depend on one another. Clarifying the constraint between these terms is therefore essential to understanding the significance of this force-field decomposition. In this section, we use the spatial representation $\gamma({\bf r}+\frac{\delta{\bf r}}{2},{\bf r}-\frac{\delta{\bf r}}{2},t)$ of the 1RDM and classify the off-diagonal terms according to their dependence on the relative coordinate $\delta{\bf r}$ to derive the hierarchy inherent in ${\bf F}_{\rm xc}^{W}$ and ${\bf F}_{\rm c}^{T}$. 

We now ask whether a current not captured by an approximate reference wavefunction can arise without taking into account a kinetic-correlation force. For the mathematical proof, we define the kinetic-energy tensor ${\bm \Pi}[\hat{\gamma}(t),{\bf r}]$ as
\begin{equation}
    \Pi_{ij}[\hat{\gamma}(t),{\bf r}]=\frac{\partial^2}{\partial\delta x_{i}\partial\delta x_{j}}\gamma\bigg({\bf r}+\frac{\delta {\bf r}}{2},{\bf r}-\frac{\delta {\bf r}}{2},t\bigg)\bigg|_{\delta{\bf r}=0}
\end{equation}
where the three-dimensional coordinates are defined as $(x_1,x_2,x_3)={\bf r}$. We also define the vector associated with the $j$th column of the rank-two kinetic-energy tensor as ${}^t{\bf K}_j[\hat{\gamma}(t)]=(\Pi_{1j},\Pi_{2j},\Pi_{3j})[\hat{\gamma}(t)]$.

\begin{theorem}
Let the reference initial state be specified at time $t=t_0$, and define the difference between the 1RDM obtained from a reference solution using a self-consistent approximation, $\hat{\gamma}_{\rm 0}(t)$, and that obtained from the exact Schr{\"o}dinger equation, $\hat{\gamma}(t)$, as $\Delta\hat{\gamma}(t)$ satisfying the initial condition $\hat{\gamma}(t_0)=\hat{\gamma}(t)$. We also fix the transverse freedom of the kinetic-energy tensor by requiring its rotational component to be the same in the exact and reference dynamics:
\begin{equation}
    \nabla\times{\bf K}_j[\hat{\gamma}(t),{\bf r}]=\nabla\times{\bf K}_j[\hat{\gamma}_0(t),{\bf r}].~\label{eq:Kconstraint}
\end{equation}
The conditions that the first-order term in the expansion of $\Delta\gamma({\bf r}+\frac{\delta{\bf r}}{2},{\bf r}-\frac{\delta{\bf r}}{2},t)$ is nonzero and the second-order term vanishes are necessary and sufficient for the following relations:
\begin{align}
{\bf j}[\hat{\gamma}(t),{\bf r}]=\frac{1}{2i}(\nabla-\nabla')\gamma({\bf r},{\bf r'},t)|_{\bf r=r'}\not={\bf j}[\hat{\gamma}_{\rm 0}(t),{\bf r}]\label{eq:j_theorem},
\end{align}
 and
\begin{equation}
n[\hat{\gamma}(t),{\bf r}]{\bf F}_{\rm c}^{T}[\hat{\gamma}(t),{\bf r}]=n[\hat{\gamma}_0(t),{\bf r}]{\bf F}_{\rm c}^{T}[\hat{\gamma}_0(t),{\bf r}].\label{eq:vcT_theorem}
\end{equation}
Furthermore, under these conditions, there is a time interval during which the electron-interaction exchange-correlation force field differs from its reference value:
\begin{equation}
{\bf F}_{\rm xc}^{W}({\bf r},t)\not={\bf F}_{\rm xc,0}^{W}({\bf r},t)
,\label{eq:wxc_theorem}
\end{equation}
where the subscript ``0'' denotes the reference exchange-correlation force field.
\end{theorem}
The proof is provided in the Appendix~\ref{appendix:proof}. Equation (10) does not imply that the rotational component of the kinetic-energy tensor is generally independent of the 1RDM. Rather, at a fixed time \(t\), inspection of the equation of motion for the current density shows that this rotational component does not enter as an independent source term for the instantaneous time derivative \(\partial_t\mathbf j\)~\cite{PhysRevLett.98.025001}. Its contribution is therefore not resolved by the local force-balance equation at that instant. We accordingly fix this transverse freedom by requiring the rotational components of \(\mathbf K_j\) to coincide between the exact and reference dynamics. Thus, the condition in Eq.~\eqref{eq:Kconstraint} fixes the transverse freedom of the second-order 1RDM tensor, which is not determined by the density force-balance equation alone.

We now discuss the physical picture implied by the mathematical constraint between ${\bf F}_{\rm xc}^{W}$ and ${\bf F}_{\rm c}^{T}$. When electron correlation drives the electron density away from its reference distribution, the resulting time-dependent density necessarily gives rise to a current that is not captured by the reference 1RDM. There is a regime in which a current beyond that of the reference state can be generated by the electron-interaction exchange-correlation force alone, without a kinetic-correlation force. The theorem identifies a regime in which the interaction component alone can generate a current beyond the reference state while the kinetic component remains representable within the reference ansatz. This regime is characterized by a nonzero first-order off-diagonal term in the 1RDM. When quadratic and higher-order off-diagonal terms become relevant, the kinetic-correlation component can no longer be represented by its reference value. 

For example, truncating the 1RDM expansion at first order in the relative coordinate $\delta{\bf r}$ yields a picture in which quantum correlations associated with spatial nonlocality are relatively weak. This picture is also consistent with the intuition behind the adiabatic approximation within the reference space, which is rooted in constructing functionals from spatially local quantities such as the density. By contrast, in systems involving ultrafast many-electron dynamics or ionization, kinetic correlation is generally important for describing nonstationary currents; capturing these currents requires the quadratic term in the relative-coordinate expansion of the 1RDM. Although the qualitative connection between spatial nonlocality and kinetic correlation is physically intuitive, the present result makes this connection explicit at the level of the exact force-balance equation. The linear and quadratic orders in the off-diagonal expansion of the 1RDM enter different components of the time-dependent exchange-correlation force. 

Finally, we discuss the classes of equations of motion distinguished by this theorem and formulated in terms of exchange-correlation force fields. Approximations to the equation of motion of the density field in Eq.~\eqref{eq:densityEOM} can be divided into three main classes. Class (i) treats ${\bf F}_{\rm xc}^{W}$ approximately and takes ${\bf F}_{\rm c}^{T}$ to be a trivial solution. Class (ii) treats ${\bf F}_{\rm xc}^{W}$ with high accuracy while still taking ${\bf F}_{\rm c}^{T}$ to be a trivial solution. Class (iii) is the exact equation of motion, in which both force fields are treated exactly.

\section{Utility of the spatial-order hierarchy}
\label{sec:utility}

The spatial-order hierarchy established in Theorem~1 can be
used not only to classify the exact exchange-correlation
force field, but also to diagnose which component of an
approximate exchange-correlation force field requires a
nontrivial correction.
Because the reference 1RDM $\hat{\gamma}_0(t)$ in
Theorem~1 is not restricted to a particular mean-field
approximation, the reference space can be chosen according
to the physical problem and the approximation to be tested.
In particular, Eqs.~(\ref{eq:j_theorem})--%
(\ref{eq:wxc_theorem}) distinguish two qualitatively
different situations.
When the deviation
$\Delta\hat{\gamma}(t)=\hat{\gamma}(t)-\hat{\gamma}_0(t)$
contains a nontrivial linear off-diagonal contribution while
the quadratic contribution remains representable within the
reference space, the current may depart from the reference
dynamics without requiring a nontrivial correction to the
kinetic-correlation component ${\bf F}_{\rm c}^{T}$. In this regime, the essential correction is carried by the
electron-interaction component ${\bf F}_{\rm xc}^{W}$. Once the quadratic off-diagonal structure becomes nontrivial, however, ${\bf F}_{\rm c}^{T}$ must also go beyond the reference representation.

The hierarchy therefore provides a criterion for determining
whether an improvement of ${\bf F}_{\rm xc}^{W}$ alone can be
sufficient or whether the approximation space must be
enlarged to include a nontrivial kinetic-correlation
contribution. This distinction is particularly useful because the two
components require qualitatively different many-body
information.
The interaction component is determined by the
pair-correlation function
$g_{\rm xc}(\mathbf r,\mathbf r',t)$, whereas ${\bf F}_{\rm c}^{T}$ depends explicitly on the off-diagonal
structure of the interacting 1RDM.
Consequently, identifying a regime in which the
kinetic-correlation contribution remains representable
within a chosen reference space narrows the many-body
information that must be improved.

The theorem should therefore not be interpreted as a closed
prescription for constructing a density-only functional.
Rather, it provides a component-selection and consistency
criterion for constructing or assessing nonadiabatic
approximations.
Below we discuss three possible uses of this criterion.

\subsection{Diagnostic criterion for adiabatic approximations in TDDFT}

A first application is the validation and possible improvement of an adiabatic approximation imposed in TDDFT.
In the adiabatic approximation of TDDFT, we approximate the exchange-correlation potential $v_{\rm xc}$ as the density functional derivative of the ground-state exchange-correlation energy $E_{\rm xc,gs}$, that is $v_{\rm xc}(t)=\delta E_{\rm xc,gs}[n(t)]/\delta n(t)$.
Let $\hat{\gamma}_{\rm ad-DFT}(t)$ denote a reference 1RDM associated with the adiabatic approximation, and suppose that the corresponding pair-correlation information required to construct ${\bf F}_{\rm xc}^{W}$ is available either explicitly or through an approximate reconstruction.
We then choose
\begin{align}
 \hat{\gamma}_0(t)=\hat{\gamma}_{\rm ad-DFT}(t),
\end{align}
in Theorem~1 and compare the resulting reference dynamics
with an exact or high-level correlated dynamics. Note that the 1RDM $\hat{\gamma}_{\rm ad-DFT}(t)$ is not the KS-1RDM but the 1RDM of the realistic system. 

Although computing the 1RDM in the realistic system directly from a DFT calculation is demanding, we need not compute the 1RDM itself; instead, we can obtain the interaction component ${\bf F}_{\rm xc}^{W}$ via the exchange-correlation kernel defined as $f_{\rm xc}(t)=\frac{\delta^2 E_{\rm ex, gs}[n(t)]}{\delta n(t)\delta n(t)}$, which gives the pair-correlation function $g_{\rm xc}(t)$. For the kinetic component, we define the longitudinal component of ${\bf F}_{\rm c}^T$ as $\nabla\cdot{\bf F}_{\rm c,\parallel}^T=\nabla v_{\rm xc}^2-\nabla\cdot{\bf F}_{\rm xc}^{W}$. We consider only the longitudinal force fields because Eq.~\eqref{eq:densityEOMKS} describes only the longitudinal component of the current density, and the description of the transverse field is outside the scope of this subsection.

If Eq.~(\ref{eq:vcT_theorem}) and Eq.~(\ref{eq:wxc_theorem}) remain satisfied, the
kinetic-correlation force is still representable within the adiabatic reference space, whereas the interaction component is not. In this case, the dominant deficiency of the approximation
can be attributed to ${\bf F}_{\rm xc}^{W}$, and an improvement of the interaction part may be sufficient without introducing an additional nonadiabatic structure into ${\bf F}_{\rm c}^{T}$.
In terms of the 1RDM, the relevant second-order off-diagonal information entering the kinetic-correlation
force is already represented at the reference level, whereas
the first-order sector responsible for the current requires
a correction.

By contrast, a violation of Eq.~(\ref{eq:vcT_theorem}) indicates that the
quadratic off-diagonal structure has left the adiabatic reference space.
In such a case, improving the interaction component alone
cannot provide a complete correction to the exact force
balance, and an explicit nonadiabatic correction to
${\bf F}_{\rm c}^{T}$ becomes necessary. The spatial-order hierarchy can therefore serve as a
criterion for distinguishing errors that may be corrected within an interaction-only description from those that signal a breakdown of the adiabatic representation itself.

\subsection{Benchmarking against high-level many-body dynamics}

A second application is to use a correlated many-body method
as a reference for the development and assessment of time-dependent density functionals. For systems in which an accurate wave function, reduced
density matrix functional~\cite{DMFT1,DMFT2}, or Green's function~\cite{PhysRev.139.A796,PhysRev.127.1391,PhysRev.124.287,PhysRev.118.1417} is available, we can instead regard $\hat{\gamma}(t)$ as the solution obtained using one of these methods. The corresponding density and current can be used to determine the effective Kohn--Sham potential, for example through an
inverse Kohn--Sham construction~\cite{TKatoiKS,PhysRevA.46.2337,PhysRevA.47.R1591,PhysRevA.49.2421}.
The accurate many-body calculation then provides the 1RDM and pair-correlation function,
from which the two components ${\bf F}_{\rm xc}^{W}$ and ${\bf F}_{\rm c}^{T}$ can be evaluated.

Given a candidate TDDFT approximation whose 1RDM in the realistic system is regarded as $\hat{\gamma}_0(t)$, one can compare its
1RDM representation with the high-level reference by
applying Eqs.~(\ref{eq:j_theorem})--(\ref{eq:wxc_theorem}).
If the kinetic-correlation condition is accurately satisfied
but the current or the interaction component is not, the
hierarchy identifies ${\bf F}_{\rm xc}^{W}$ as the primary target
for functional improvement.
Conversely, when the quadratic off-diagonal condition is no
longer satisfied, the discrepancy cannot be removed solely
by correcting ${\bf F}_{\rm xc}^{W}$, and the kinetic-correlation
component must also be improved.

This provides a component-resolved benchmarking strategy.
Instead of judging an approximate functional only based on the
total density or the total exchange-correlation potential,
the spatial-order hierarchy identifies the lowest
off-diagonal order at which the approximate representation
departs from the high-level reference.

\subsection{Reference-space diagnosis of nonadiabatic dynamics}

The freedom to choose $\hat{\gamma}_0(t)$ also allows the
hierarchy to be used as a more general diagnostic of
nonadiabatic dynamics.
As an example, consider electron dynamics coupled to nuclear
motion by introducing the time-dependent nuclear coordinate ${\bf R}\rightarrow{\bf R}(t)$.
For each nuclear configuration, an adiabatic electronic
state defines a reference 1RDM $\hat{\gamma}_{\rm adia}(t)$.
The difference
\begin{align}
 \Delta\hat{\gamma}_{\rm nad}(t)
 =
 \hat{\gamma}(t)-\hat{\gamma}_{\rm adia}(t)
\end{align}
then characterizes the departure of the actual electronic
motion from the adiabatic reference.

In the slow-nuclear-dynamics limit, the second time derivative of the difference between the 1RDMs is approximately given by
\begin{equation}
    -\frac{\partial^2}{\partial t^2}\Delta\gamma({\bf r},{\bf r'},t)\sim \bigg[i\frac{\partial}{\partial t}(v_{\rm ext}({\bf r},t)-v_{\rm ext}({\bf r}',t))\bigg]\gamma_{\rm adia}({\bf r},{\bf r'},t).\label{eq:deltagamma_2ndderiv}
\end{equation}
The derivation of Eq.~\eqref{eq:deltagamma_2ndderiv} is summarized in Appendix B.

Under the adiabatic assumption, the 1RDM $\hat{\gamma}_{\rm adia}(t)$ can be made real-valued by applying an appropriate gauge transformation, owing to the steady-state character of the reference. In this case, the 1RDM is even under exchange of the spatial coordinates, so that $\gamma_{\rm adia}({\bf r},{\bf r'},t)=\gamma_{\rm adia}({\bf r}',{\bf r},t)$, since the Hermiticity of the 1RDM must be satisfied. By contrast, the force term in Eq.~\eqref{eq:deltagamma_2ndderiv} is odd under exchange of the spatial coordinates. These conditions, therefore, imply that the 1RDM $\Delta\gamma({\bf r},{\bf r'},t)$ must be odd under exchange of the spatial coordinates.

The odd parity of $\Delta\gamma({\bf r},{\bf r'},t)$ leads to a non-zero linear term of the relative coordinate $\delta{\bf r}={\bf r}-{\bf r}'$ associated
with the current, whereas the quadratic contribution relevant to the kinetic-correlation force vanishes.
A non-zero linear term therefore identifies an electronic
current that cannot be represented within the instantaneous
adiabatic reference.
If Eq.~(\ref{eq:vcT_theorem}) nevertheless remains
satisfied, this departure is still confined to the
first-order spatial sector.
When Eq.~(\ref{eq:vcT_theorem}) is violated, the
nonadiabatic correction has propagated into the quadratic
sector of the 1RDM, implying that a modification of the
kinetic-correlation force is also required.

This observation suggests that the spatial-order hierarchy
can be used to distinguish different levels of departure
from a chosen adiabatic or approximate reference manifold.
The same construction is not restricted to density-functional approximations. The hierarchy thus provides a common language for comparing
different many-body representations according to the
spatial order at which they cease to reproduce the exact
dynamics.
In electron--nuclear problems, such a classification may
also provide a basis for constructing variational or
density-functional approximations that incorporate
vibronic and nonadiabatic corrections in a controlled
hierarchical manner.

\section{Numerical calculation: one-dimensional model}
We numerically demonstrate the electron and force field dynamics under the one-dimensional modeling; note that the force field decomposition itself is not restricted to one dimension. The spatial-order criterion derived in Sec.IV is formulated for a three-dimensional \(N\)-electron system; the one-dimensional two-electron model employed below serves only as a numerical illustration. For a one-dimensional two-electron spin-singlet system, the TDKS spatial orbital is determined as $\phi(x,t)=\sqrt{n(x,t)/2}\exp\left[i\int^x dy\,u(y,t)\right]$~\cite{KinC-XC-JCP2014,Bauer2006}, and its 1RDM is
\begin{equation}
 \gamma_s(x,x',t)=\sqrt{n(x,t)n(x',t)}
 \exp\!\left[i\int_x^{x'}d\xi\,u(\xi,t)\right],
 \label{eq:gammas}
\end{equation}
which is a rank-one 1RDM. Therefore, using the obtained $n$ and $u$, identifying $\hat{\gamma}_{\rm HF}(t)$ with
$\hat{\gamma}_{s}(t)$ gives the trivial solution $F_{\rm c}^{T}=0$, the trivial solution of $F_{\rm c}^{T}$ in the approximate equation classes is $F_{\rm c}^{T}=0$ from Eq.~\eqref{eq:vcT_theorem}.

We therefore classify the equation classes according to the presence or absence of truncation by $F_{\rm c}^{T}=0$, and visualize and discuss how this affects electron-density motion. In what follows, we performed a numerical analysis of a two-electron system modeled after a one-dimensional helium model. Here, for a problem setting in which the system was in the HF ground state at $t<0$ and was switched to the bare two-electron interaction at $t=0$, we examined how the electron-density motion behaves differently among the following three classified equation classes: (i) the TDHF approximation; (ii) $F_{\rm xc}^{W}$ constructed from the exact wavefunction determined by the two-electron TDSE, with $F_{\rm c}^{T}=0$; and (iii) the exact $F_{\rm xc}^{W}$ and $F_{\rm c}^{T}$ determined by the two-electron TDSE. In the TDHF approximation, the exchange-correlation potential under the HF approximation is represented by $-\frac{1}{2}v_{\rm H}(x,t)$, corresponding to one half of the Hartree potential.

In the numerical calculation, the external potential $v_{\rm ext}$ and the electron-electron potential $w_{\rm ee}$ are given as follows: $v_{\rm ext}(x,t)=-\frac{2}{\sqrt{1+x^2}}$
and
$w_{\rm ee}(x,y)=\frac{1}{\sqrt{1+(x-y)^2}}$, corresponding to the soft-Coulomb interaction model. For the initial state at $t=0$, the numerical solution of the HF approximation $\Psi(x_1,x_2,0)=\varphi_{\rm HF}(x_1)\varphi_{\rm HF}(x_2)$ was used. This numerical solution was optimized by the imaginary-time propagation method.

The time evolution was carried out by a sequential product of operators based on the Suzuki-Trotter decomposition~\cite{Trotter1959On,Suzuki1976Generalized}, which partitions the total Hamiltonian into the kinetic energy operator and the potential operator. Here, the time-step width $\Delta t$ satisfies the relation $t=n\Delta t$; $\Delta t=0.01$ a.u. was used for time propagation. Furthermore, acting the kinetic energy operator on the wavefunction and the KS orbitals was treated as a local operation in wave-number space by discrete Fourier transforms of the wavefunction and operator. The wavefunction and KS orbitals were represented on a grid with spacing $\Delta x=0.1$ a.u., and the domain was set to $x\in[-15.0,15.0]$. The total simulation time $T$ was set to 20.0 a.u.
\section{Time dependence of the density}
\begin{figure}[t]
    \centering
    \includegraphics[width=\columnwidth]{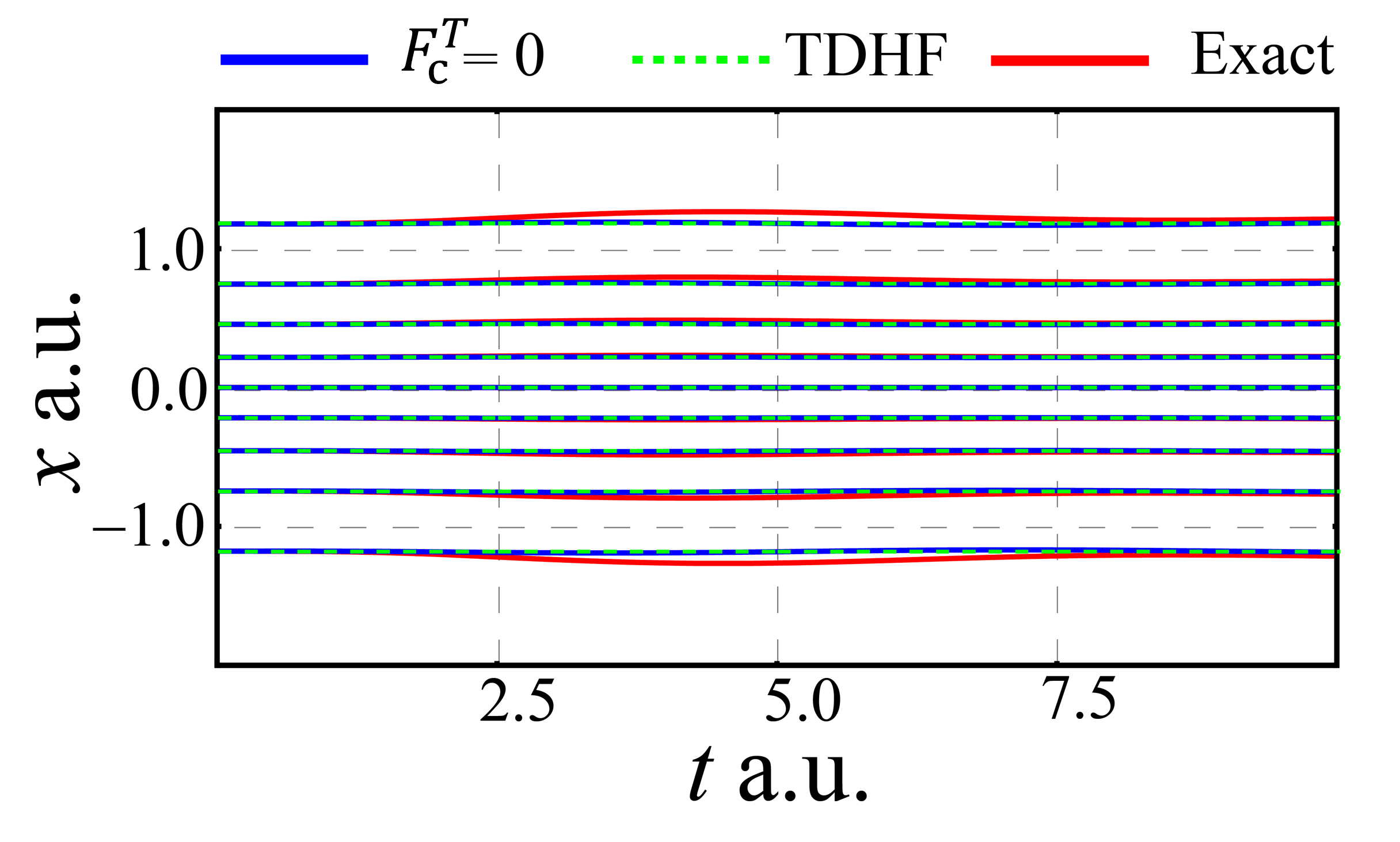}
    \caption{Contours of the cumulative density
$\Theta(x,t)$ obtained from the exact TDSE (red solid),
the truncated propagation with exact $F_{\rm xc}^{W}$ and
$F_{\rm c}^{T}=0$ (blue dashed), and TDHF (green dotted).
The contour interval is 0.2.}
    \label{fig:Veff300}
\end{figure}

In this section, we numerically verify that $F_{\rm c}^{T}$ characterizes the exact class (iii) and actually produces a decisive difference from the approximate classes (i) and (ii). First, we compare the exact electron density with the electron density obtained from the approximation of $F_{\rm c}^{T}=0$. To make the difference clear, the cumulative distribution function, which is an integrated form of the electron density, $\Theta(x,t)=\int^{x}_{-L}dx'n(x',t)$, is shown as contour lines in Fig.~\ref{fig:Veff300}. The electron density is denser in regions where the contour lines are more closely spaced.
Focusing on the deviation from the TDHF solution (green dashed lines in Fig.~\ref{fig:Veff300}), it can be seen that the contour lines of the exact solution exhibit delocalization until $t=5.0$ a.u. away from the nuclear center $x=0.0$ a.u. At time $t=5.0$ a.u., the spacing between the contour lines of the exact solution is the widest. After passing $t=5.0$ a.u., they converge toward the TDHF solution until $t=10.0$ a.u., although the contour lines do not completely coincide. Such behavior of the electron density is not observed at all in the one-electron motion when $F_{\rm c}^{T}=0$ is imposed as a truncation, and no difference from the TDHF solution is seen.

\section{One-electron effective force field analysis}
To discuss how the presence or absence of truncation of $F_{\rm c}^{T}$ produces an essential difference in electron-density motion, we define the one-electron effective force field describing the equation of motion of the electron density from Eq.~\eqref{eq:densityEOM} or~\eqref{eq:densityEOMKS}. In the hydrodynamic framework, the equation for the Lagrangian derivative of the particle velocity field $u(x,t)=j(x,t)/n(x,t)$ is derived as a simultaneous solution of the mass-conservation law and momentum-conservation law for the density field:
\begin{equation}
    \partial_tu(x,t)+u(x,t)\partial_xu(x,t)=F_{\rm eff}(x,t).
\end{equation}
In this study, the source term $F_{\rm eff}(x,t)$ on the right-hand side is called the effective force field. If the effective force field is known, the velocity field $u(x,t)$ is obtained, and, from the mass-conservation law $\partial_tn=-\partial_x(n u)$, the motion of the electron density $n$ is determined. Rearranging Eq.~\eqref{eq:densityEOMKS}, since KS-1RDM $\hat{\gamma}_s(t)$ is determined as Eq.~\eqref{eq:gammas}~\cite{KinC-XC-JCP2014,Bauer2006}, the form of $F_{\rm eff}(x,t)$ in one-dimensional two-electron system is simply given by
\begin{equation}
F_{\rm eff}(x,t)=-\partial_x\left(v_{\rm ext}+v_{\rm H}+v_{\rm xc}-\frac{1}{2\sqrt{n}}\partial_x^2\sqrt{n}\right)\label{eq:Feff2}.
\end{equation}
Thus, if the electron density obtained from the approximate equation class (ii) agrees with the exact electron density at any time, the difference between the approximated and exact effective force fields $F_{\rm eff}$ is necessarily equal to $F_{\rm c}^T$.

\begin{figure}[t]
    \centering
    \includegraphics[width=1.0\columnwidth]{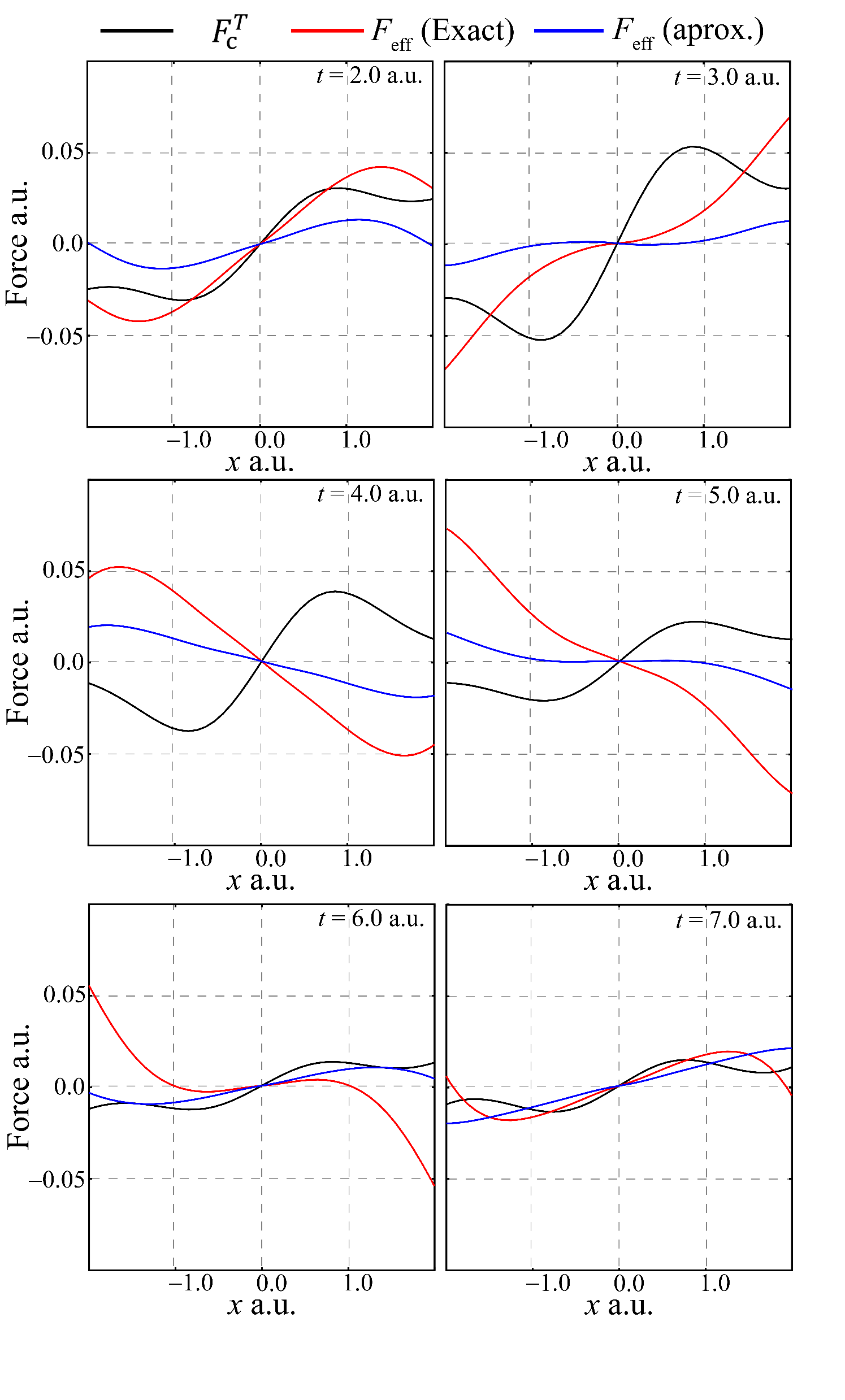}
    \caption{Snapshots of kinetic-correlation force $-\partial_xv_c^T$ (black solid), the exact effective force $F_{\rm eff}^{\rm exact}$ (red dashed), and the effective force obtained with $F_{\rm c}^{T}=0$ (blue dotted).}
    \label{fig:Feff}
\end{figure}
Figure~\ref{fig:Feff} shows that $F_c^T$ points away from the nuclear center throughout the propagation and therefore drives the initial density broadening. The exact $F_{\rm eff}$ reverses sign at $t\sim4.0$ a.u. as the remaining attractive contributions become dominant and reverses again near $t\sim7.0$ a.u., producing the oscillatory density motion. In contrast, the truncated effective force remains too weak near the density maximum to generate the observed delocalization.

Finally, it should be noted that the exact $F_{\rm c}^{T}$ obtained from the TDSE does not agree with the difference between the effective force fields obtained from the exact calculation and the truncated approximate calculation with $F_{\rm c}^{T}=0$. As stated above, this corresponds to numerically verifying the treatment of the proposition that, if the electron densities agree at any time, the difference between their respective effective force fields $F_{\rm eff}$ is necessarily equal to $F_{\rm c}^T$.

\section{Conclusion}
We have extended the decomposition of the exchange-correlation potential into interaction and kinetic components, originally introduced in Ref.~\cite{KinC-XC-JCP2014}, to three-dimensional systems. In the present study, we instead formulated the two components in terms of force fields to account for the nontrivial rotational components that drive the rotational current density.
Following this decomposition, we proved a mathematical constraint relating the interaction
and kinetic components of the time-dependent exchange-correlation
potential. A linear off-diagonal deviation of the exact 1RDM from
that of an approximate reference can generate a current not captured by the reference approximation while the kinetic component remains representable, whereas quadratic and
higher-order off-diagonal structures activate kinetic correlation. In a
two-electron correlation quench, retaining the exact interaction
component while setting $F_{\rm c}^{T}=0$ suppresses the density broadening
observed in the exact evolution. These results show that the two
components are not independent approximation targets and provide a
structural constraint for constructing nonadiabatic functionals.

Although this work provides a framework for classification, the
classification criterion is relatively simple because we consider only
the two components of the exchange-correlation potential and define
triviality with respect to the reference representation space. Identifying
the mathematical constraints among more than two partitioned terms and
classifying them with respect to representation spaces would require a more complex analysis. Clarifying the hierarchical structure of the exchange-correlation potential in greater detail may lead to a framework for constructing functionals that more closely reflects the desired physical picture.

\section{Acknowledgement}
This work was supported by the Japan Society for the Promotion of
Science (JSPS) through the Overseas Research Fellowships program.
\bibliography{TDNO_EPT}

\appendix
\onecolumngrid
\setcounter{equation}{0}
\renewcommand{\theequation}{A\arabic{equation}}

\section{Proof of the Theorem \label{appendix:proof}}
For $t>t_0$, the difference between the 1RDMs, $\Delta \hat{\gamma}(t)=\hat{\gamma}(t)-\hat{\gamma}_{\rm 0}(t)$, can be expanded in terms of the relative coordinate $\delta {\bf r}={\bf r}-{\bf r}'$ and the centroid coordinate ${\bf r}_{\rm G}=\frac{{\bf r}+{\bf r}'}{2}$ as follows:
\begin{equation}
\begin{split}
    \gamma({\bf r},{\bf r'},t)&=\gamma_{\rm 0}({\bf r},{\bf r'},t)+\Delta\gamma\bigg({\bf r}_{\rm G}+\frac{\delta {\bf r}}{2},{\bf r}_{\rm G}-\frac{\delta {\bf r}}{2},t\bigg)\\
    &=\gamma_{\rm 0}({\bf r},{\bf r'},t)+\Delta\gamma\bigg({\bf r}_{\rm G},{\bf r}_{\rm G},t\bigg)+\frac{\delta {\bf r}}{2}\cdot\bigg(\nabla\Delta\gamma\big({\bf r},{\bf r}',t\big)\big|_{\bf r=r'=r_{\rm G}}-\nabla'\Delta\gamma\big({\bf r},{\bf r}',t\big)\big|_{\bf r=r'=r_{\rm G}}\bigg)\\
    &+\frac{1}{2}\bigg(\frac{\delta{\bf r}\cdot(\nabla-\nabla')}{2}\bigg)^2\Delta\gamma({\bf r},{\bf r'},t)|_{{\bf r}={\bf r'}={\bf r}_{\rm G}}+O(|{\delta}{\bf r}|^3).
\end{split}
\end{equation}
The density and current density are defined in terms of the 1RDM as $n[\hat{\gamma}(t),{\bf r}]={\gamma}({\bf r},{\bf r},t)$ and ${\bf j}[\hat{\gamma}(t),{\bf r}]=\frac{1}{2i}(\nabla-\nabla'){\gamma}({\bf r},{\bf r'},t)\big|_{\bf r=r'}$, respectively. Applying these definitions to $\Delta \hat{\gamma}(t)$ allows us to rewrite the expansion above as
\begin{equation}
\begin{split}
    \gamma({\bf r},{\bf r'},t)&=\gamma_{\rm 0}({\bf r},{\bf r'},t)+n\big[\Delta \hat{\gamma}(t),{\bf r}_{\rm G}\big]+i\delta{\bf r}\cdot{\bf j}\big[\Delta\hat{\gamma}(t),{\bf r}_{\rm G}\big]+\frac{1}{2}\bigg(\frac{{\delta}{\bf r}\cdot(\nabla-\nabla')}{2}\bigg)^2\Delta\gamma({\bf r},{\bf r'},t)|_{{\bf r}={\bf r'}={\bf r}_{\rm G}}+O(|\delta{\bf r}|^3).\label{eq:gamma_expansion}
\end{split}
\end{equation}

We first prove the sufficient condition: if the 1RDM $\Delta\gamma({\bf r}+\delta{\bf r},{\bf r}-\delta{\bf r},t)$ has a nonzero first-order term and a vanishing second-order term in its expansion in $\delta{\bf r}$, then Eqs.~\eqref{eq:j_theorem} and~\eqref{eq:vcT_theorem} follow. The Taylor expansion of Eq.~\eqref{eq:gamma_expansion} therefore has a nonzero first-order coefficient and a zero second-order coefficient, respectively. Thus, we obtain
\begin{equation}
    {\bf j}[\Delta \hat{\gamma}(t),{\bf r}_{\rm G}]\not=0,\quad{\rm and}\quad \frac{1}{4}\bigg(\frac{\partial}{\partial x_i}-\frac{\partial}{\partial x_i'}\bigg)\bigg(\frac{\partial}{\partial x_j}-\frac{\partial}{\partial x_j'}\bigg)\Delta\gamma({\bf r},{\bf r'},t)|_{{\bf r}={\bf r'}={\bf r}_{\rm G}}=0,\quad{{\rm for\,\,any}\,\,i,j\in\{1,2,3\}}\label{eq:sufficientcond}
\end{equation}
where $(x_1,x_2,x_3)={\bf r}$. The left-hand relation is equivalent to ${\bf j}\big[\hat{\gamma}(t)\big]\not={\bf j}\big[\hat{\gamma}_{\rm 0}(t)\big]$. For the right-hand relation, taking derivatives with respect to the centroid coordinate, represented by $\frac{\partial}{\partial x_i}+\frac{\partial}{\partial x_i'}$, and summing over the directions $x_i\in\{x,y,z\}$, yields
$n[\hat{\gamma}(t),{\bf r}]{\bf F}_{\rm c}^T[\hat{\gamma}(t),{\bf r}]=n[\hat{\gamma}_0(t),{\bf r}]{\bf F}_{\rm c}^T[\hat{\gamma}_{\rm 0}(t),{\bf r}]$.

We next prove the necessary condition. The inequality ${\bf j}\big[\hat{\gamma}(t)\big]\not={\bf j}\big[\hat{\gamma}_{\rm 0}(t)\big]$ immediately implies that the first-order coefficient is nonzero. From the definition in Eq.~\eqref{eq:vcT}, $n\nabla v_{\rm c}^{T}$ corresponds to the second-order coefficient in $\delta {\bf r}$. Therefore, the assumption
$n[\hat{\gamma}(t),{\bf r}]{\bf F}_{\rm c}^T[\hat{\gamma}(t),{\bf r}]=n[\hat{\gamma}_0(t),{\bf r}]{\bf F}_{\rm c}^T[\hat{\gamma}_{\rm 0}(t),{\bf r}]$
implies
\begin{equation}
    \sum_{i=1}^3\frac{\partial}{\partial x_{{\rm G},i}}\frac{\partial^2}{\partial\delta x_{i}\partial\delta x_{j}}\gamma\bigg({\bf r}+\frac{\delta {\bf r}}{2},{\bf r}-\frac{\delta {\bf r}}{2},t\bigg)\bigg|_{\delta{\bf r}=0}=\sum_{i=1}^3\frac{\partial}{\partial x_{{\rm G},i}}\frac{\partial^2}{\partial\delta x_{i}\partial\delta x_{j}}\gamma_{\rm 0}\bigg({\bf r}+\frac{\delta {\bf r}}{2},{\bf r}-\frac{\delta {\bf r}}{2},t\bigg)\bigg|_{\delta{\bf r}=0},\label{eq:kinetic_tensor_assumption}
\end{equation}
for every $j$. By Helmholtz's theorem, this relation implies
\begin{equation}
    \frac{\partial^2}{\partial\delta x_{i}\partial\delta x_{j}}\gamma\bigg({\bf r}+\frac{\delta {\bf r}}{2},{\bf r}-\frac{\delta {\bf r}}{2},t\bigg)\bigg|_{\delta{\bf r}=0}=\frac{\partial^2}{\partial\delta x_{i}\partial\delta x_{j}}\gamma_{\rm 0}\bigg({\bf r}+\frac{\delta {\bf r}}{2},{\bf r}-\frac{\delta {\bf r}}{2},t\bigg)\bigg|_{\delta{\bf r}=0}+\left[\nabla\times {\bf A}_j({\bf r},t)\right]_{i},\label{eq:gauge}
\end{equation}
where $\left[\nabla\times {\bf A}_j({\bf r},t)\right]_{i}$ denotes the $i$-th component of the rotational vector field $\nabla\times {\bf A}_j({\bf r},t)$, and the vector field ${\bf A}_j({\bf r},t)$ arises from the Helmholtz decomposition.

Furthermore, we impose the condition that the rotational parts of the exact and reference kinetic-energy tensors coincide. Specifically, we write $\Pi_{ij}[\hat{\gamma}(t),{\bf r}]=\frac{\partial^2}{\partial\delta x_i\partial\delta x_j}\gamma({\bf r}+\frac{\delta {\bf r}}{2},{\bf r}-\frac{\delta {\bf r}}{2},t)|_{\delta{\bf r}=0}=[{\bf K}_j[\hat{\gamma}(t),{\bf r}]]_i$, which gives $\nabla\times{\bf K}_i[\hat{\gamma}(t),{\bf r}]=\nabla\times{\bf K}_i[\hat{\gamma}_0(t),{\bf r}]$. This leads to the relation 
\begin{equation}
    \nabla\times\frac{\partial}{\partial\delta {\bf r}}\frac{\partial}{\partial\delta x_{j}}\gamma\bigg({\bf r}+\frac{\delta {\bf r}}{2},{\bf r}-\frac{\delta {\bf r}}{2},t\bigg)\bigg|_{\delta{\bf r}=0}=\nabla\times\frac{\partial}{\partial\delta {\bf r}}\frac{\partial}{\partial\delta x_{j}}\gamma_{0}\bigg({\bf r}+\frac{\delta {\bf r}}{2},{\bf r}-\frac{\delta {\bf r}}{2},t\bigg)\bigg|_{\delta{\bf r}=0}
\end{equation}
for any $j$. This relation therefore implies the following condition for ${\bf A}_j({\bf r},t)$:
\begin{equation}
    \nabla\times\big({\nabla}\times{\bf A}_j({\bf r},t)\big)=0,
\end{equation}
Applying Helmholtz's theorem once more to the rotational field ${\nabla}\times{\bf A}_j({\bf r},t)$, we obtain the gradient of a scalar function, ${\nabla}\times{\bf A}_j({\bf r},t)=\nabla\phi_j({\bf r},t)$. Substituting ${\nabla}\times{\bf A}_j({\bf r},t)=\nabla\phi_j({\bf r},t)$ into Eq.~\eqref{eq:gauge}, taking the divergence with respect to the centroid coordinate, and using the assumption in Eq.~\eqref{eq:kinetic_tensor_assumption} again, we obtain the following equation:
\begin{equation}
    \nabla^2\phi_j({\bf r},t)=0.
\end{equation}
This is Laplace's equation. For open boundary conditions, we impose the condition that the scalar function is regular and approaches a finite constant at infinity, as $|{\bf r}|\rightarrow \infty$. The scalar function then has the trivial solution $\phi_j({\bf r},t)=const.$, resulting in Eq.~\eqref{eq:sufficientcond}. For periodic boundary conditions, we can take the Fourier transform of $\phi_j({\bf r},t)$, resulting in $k^2\tilde{\phi}({\bf k},t)=0$ for ${\bf k}\not=0$. This implies that $\phi_j({\bf r},t)=const.$, resulting in Eq.~\eqref{eq:sufficientcond}.

Finally, we prove the condition Eq.~\eqref{eq:wxc_theorem} concerning ${\bf F}_{\rm xc}^W$. At time $t$, the equations of motion for the respective 1RDMs are expressed as follows:
\begin{equation}
\begin{split}
    i\partial_t\gamma({\bf r},{\bf r'},t)&=\bigl(\hat{h}({\bf r},t)-\hat{h}({\bf r'},t)\bigr)\gamma({\bf r},{\bf r'},t)+\int d{\bf r''}\bigl(w_{\rm ee}(|{\bf r}-{\bf r''}|)-w_{\rm ee}(|{\bf r'}-{\bf r''}|)\bigr)\Gamma({\bf r}{\bf r''}|{\bf r'}{\bf r''},t).
\end{split}
\end{equation}
The operator $\hat{h}({\bf r},t)$ is the one-electron Hamiltonian, $\hat{h}({\bf r},t)=-\frac{1}{2}\nabla^2+v_{\rm ext}({\bf r},t)$. The two-electron reduced density matrix $\Gamma({\bf r}{\bf r''}|{\bf r'}{\bf r''},t)$ is defined as follows:
\begin{equation}
\begin{split}
    \Gamma({\bf r}{\bf r''}|{\bf r'}{\bf r''},t)=N(N-1)&\int d\sigma_1\cdots d\sigma_N\int d{\bf r}_3\cdots d{\bf r}_N\Psi({\bf r}\sigma_1,{\bf r''}\sigma_2,\cdots{\bf r}_N\sigma_N,t)\Psi^*({\bf r'}\sigma_1,{\bf r''}\sigma_2,\cdots{\bf r}_N\sigma_N,t).
\end{split}
\end{equation}
In addition, at $t=t_0$, the boundary condition $\hat{\gamma}_{\rm 0}(t_0)=\hat{\gamma}(t_0)$ holds. If we regard the reference state as the result obtained by some approximation, the 2RDM is approximated as $\Gamma({\bf r}{\bf r''}|{\bf r'}{\bf r''},t)=\Gamma_0({\bf r}{\bf r''}|{\bf r'}{\bf r''},t)$.

Taking the difference between the exact equation and the reference equation obtained under the approximation gives the following form:
\begin{equation}
\begin{split}
    \big(i\partial_t-h({\bf r},t)+h({\bf r'},t)\big)\Delta \gamma({\bf r},{\bf r'},t)
    &=\int d{\bf r''}\bigl(w_{\rm ee}(|{\bf r}-{\bf r''}|)-w_{\rm ee}(|{\bf r'}-{\bf r''}|)\bigr)\\
    &\times\big(\Gamma({\bf r}{\bf r''}|{\bf r'}{\bf r''},t)-\Gamma_0({\bf r}{\bf r''}|{\bf r'}{\bf r''},t))~\label{eq:1RDMEOM_exact_minus_HF}.
\end{split}
\end{equation}
The left-hand side of the above equation gives
\begin{equation}
\begin{split}
    i\partial_t\Delta\gamma({\bf r},{\bf r'},t)
    &=i\partial_tn\big[\Delta \hat{\gamma}(t),{\bf r}_{\rm G}\big]
    -\delta {\bf r}\cdot\partial_t{\bf j}\big[\Delta \hat{\gamma}(t),{\bf r}_{\rm G}\big]+O(|\delta {\bf r}|^3)\label{eq:dt_1RDM},  
\end{split}  
\end{equation}
and
\begin{equation}
\begin{split}
    \big(h({\bf r},t)-h({\bf r'},t)\big)\Delta\gamma({\bf r},{\bf r'},t)
    &=-i\nabla_{\rm G}{\bf j}\big[\Delta \hat{\gamma}(t),{\bf r}_{\rm G}\big]+\delta{\bf r}\cdot\left(\nabla_{\rm G}v({\bf r}_{\rm G},t )n\big[\Delta\hat{\gamma}(t),{\bf r}_{\rm G}\big]+v({\bf r}_{\rm G},t ){\bf j}\big[\Delta \hat{\gamma}(t),{\bf r}_{\rm G}\big]\right)+O(|\delta{\bf r}|^2).
    \label{eq:h_1RDM}
\end{split}
\end{equation}
These expressions are obtained with $\nabla_{\rm G}=\nabla+\nabla'$. The first term on the right-hand side of Eq.~\eqref{eq:dt_1RDM} and the first term on the right-hand side of Eq.~\eqref{eq:h_1RDM} cancel by virtue of the continuity equation. In addition, when taking the limit approaching the initial condition $t\rightarrow t_0+0$, $\hat{\gamma}(t)\rightarrow\hat{\gamma}(t_0)=\hat{\gamma}_{\rm 0}(t_0)$, and the terms not containing time derivatives vanish. That is,
\begin{equation}
\begin{split}
({\rm l.h.s.\,\,of\,\,Eq.}~\eqref{eq:1RDMEOM_exact_minus_HF})    &=-\delta {\bf r}\cdot\partial_t{\bf j}\big[\Delta \hat{\gamma}(t),{\bf r}_{\rm G}\big]_{t=t_0}+(t-t_0)\delta {\bf r}\cdot\partial_t^2{\bf j}\big[\Delta \hat{\gamma}(t),{\bf r}_{\rm G}\big]_{t=t_0}\\
    &+(t-t_0)\delta {\bf r}\cdot\left(\nabla_{\rm G}v({\bf r}_{\rm G},t )\partial_tn\big[\Delta \hat{\gamma}(t),{\bf r}_{\rm G}\big]+v({\bf r}_{\rm G},t )\delta{\bf r}\cdot\partial_t{\bf j}\big[\Delta \hat{\gamma}(t),{\bf r}_{\rm G}\big]\right)_{t=t_0}+\cdots+O(|\delta{\bf r}|^2) .\label{eq:dt_plus_h_1RDM}
\end{split}
\end{equation}
is obtained. Here, in Eq.~\eqref{eq:dt_plus_h_1RDM}, when the coefficient of $(t-t_0)^m$ first becomes non-zero at the lowest order $m$, the coefficient is expressed as $\partial_t^{m+1}{\bf j}\big[\Delta \hat{\gamma}(t),{\bf r}_{\rm G}\big]_{t=t_0}$. This is because it can be proved inductively from $m=0$ that $\partial_t^m(n[\hat{\gamma}(t)]-n[\hat{\gamma}_{0}(t)])=0$ and $\partial_t^m({\bf j}[\hat{\gamma}(t)]-{\bf j}[\hat{\gamma}_{0}(t)])=0$.
Similarly, the right-hand side of Eq.~\eqref{eq:1RDMEOM_exact_minus_HF} is
\begin{equation}
\begin{split}
          ({\rm r.h.s.\,\,of\,\,Eq.}~\eqref{eq:1RDMEOM_exact_minus_HF})
          &=\delta{\bf r}\cdot\int d{\bf r''}\nabla_{\rm G} w_{\rm ee}(|{\bf r}_{\rm G}-{\bf r''}|)\big(\Gamma({\bf r}_{\rm G}{\bf r''}|{\bf r}_{\rm G}{\bf r''},t_0)-\Gamma_0({\bf r}_{\rm G}{\bf r''}|{\bf r}_{\rm G}{\bf r''},t_0)\big)\\
          &+(t-t_0)\delta{\bf r}\cdot\int d{\bf r''}\nabla_{\rm G} w_{\rm ee}(|{\bf r}_{\rm G}-{\bf r''}|)\partial_t\big(\Gamma({\bf r}_{\rm G}{\bf r''}|{\bf r}_{\rm G}{\bf r''},t)-\Gamma_0({\bf r}_{\rm G}{\bf r''}|{\bf r}_{\rm G}{\bf r''},t)\big)_{t=t_0}\\
          &+\cdots+O(|\delta{\bf r}|^2).
          \label{eq:wxc_derivative}
\end{split}
\end{equation}
Similarly to Eq.~\eqref{eq:dt_plus_h_1RDM}, when the lowest order at which the coefficient becomes non-zero is $(t-t_0)^m$,
\begin{equation}
\int d{\bf r''}\nabla_{\rm G} w_{\rm ee}(|{\bf r}_{\rm G}-{\bf r''}|)\partial_t^{m}\big(\Gamma({\bf r}_{\rm G}{\bf r''}|{\bf r}_{\rm G}{\bf r''},t)-\Gamma_0({\bf r}_{\rm G}{\bf r''}|{\bf r}_{\rm G}{\bf r''},t)\big)_{t=t_0}\not=0,\label{eq:wxc_derivative2}
\end{equation}
since Eqs.~\eqref{eq:dt_plus_h_1RDM} and~\eqref{eq:wxc_derivative} are connected by equality. Considering $\partial_t^{m'}n[\hat{\gamma}(t)]|_{t=t_0}=\partial_t^{m'}n[\hat{\gamma}_{\rm 0}(t)]|_{t=t_0}$ for $m'\le m$ due to the mathematical induction in Eq.~\eqref{eq:dt_plus_h_1RDM}, we can rewrite Eq.~\eqref{eq:wxc_derivative} as
\begin{equation}
\begin{split}
    ({\rm l.h.s.\,\,of\,\,Eq.}~\eqref{eq:wxc_derivative2})&=\int d{\bf r''}\nabla_{\rm G} w_{\rm ee}(|{\bf r}_{\rm G}-{\bf r''}|)\partial_t^{m}\big(\Gamma({\bf r}_{\rm G}{\bf r''}|{\bf r}_{\rm G}{\bf r''},t)-\Gamma_0({\bf r}_{\rm G}{\bf r''}|{\bf r}_{\rm G}{\bf r''},t)\big)_{t=t_0}\\
    &=\partial_t^{m}\left(-n({\bf r}_{\rm G},t){\bf F}_{\rm xc}^{W}({\bf r}_{\rm G},t)+n_0({\bf r}_{\rm G},t){\bf F}_{{\rm xc},0}^{W}({\bf r}_{\rm G},t)\right)\\
    &=\partial_t^{m}\left(-n_0({\bf r}_{\rm G},t)\big({\bf F}_{\rm xc}^{W}({\bf r}_{\rm G},t)-{\bf F}_{{\rm xc},0}^{W}({\bf r}_{\rm G},t)\big)\right)\not=0\label{eq:2RDM_conclu}.
\end{split}
\end{equation}
This means nothing other than that, at some $t>t_0$, ${\bf F}_{\rm xc}^{W}$ departs from the reference solution. By imposing the non-triviality condition Eq.~\eqref{eq:gamma_expansion} concerning ${\bf j}$ in the above proof, the sufficient condition~\eqref{eq:2RDM_conclu} concerning ${\bf F}_{\rm xc}^{W}$ is also obtained. This proves the proposition.

Finally, as a note, we mention the residual term $O(|\delta{\bf r}|^2)$. In a case where the coefficients of $|\delta{\bf r}|^2$ in Eqs.~\eqref{eq:dt_plus_h_1RDM} and~\eqref{eq:wxc_derivative} completely agree at any time, the triviality of ${\bf F}_{\rm c}^{T}$ is guaranteed, and the theorem holds. On the other hand, they do not agree in most cases, and in that case, residual terms depending on $|\delta{\bf r}|^2$ participate in the time evolution of the 1RDM. When this occurs, ${\bf F}_{c}^{T}[\hat{\gamma}(t)]$ is no longer a trivial solution and no longer satisfies the conditions stated in the theorem.

\onecolumngrid
\setcounter{equation}{0}
\renewcommand{\theequation}{B\arabic{equation}}

\section{Derivation of Eq.~\eqref{eq:deltagamma_2ndderiv}}
The dynamics of the wavefunction is governed by the time-dependent external potential as $v_{\rm ext}({\bf r},t)=v_{\rm ext}({\bf r},{\bf R}(t))$ and approximately satisfy the eigenvalue equation of time-dependent Hamiltonian $\hat{H}({\bf r}_1,\cdots,{\bf r}_N,t)=\hat{T}+\hat{W}+\sum_{i=1}^Nv_{\rm ext}({\bf r}_i,t)$ as follows:
\begin{equation}
\begin{split}
\hat{H}(t)\ket{\Psi(t)}\sim E(t)\ket{\Psi(t)}\label{eq:adia_eigen}
\end{split}
\end{equation}
where $E(t)$ denotes eigenenergy of the ground state of the Hamiltonian at time $t$. This approximation, however, violates in considering the second order time derivative to the exact 1RDM as follows:
\begin{equation}
    \begin{split}
        -\frac{\partial^2}{\partial t^2}\gamma({\bf r},{\bf r'},t)&=i\frac{\partial}{\partial t}\int d\sigma\braket{\Psi(t)|[\hat{\psi}_{\sigma}^{\dagger}({\bf r'})\hat{\psi}_{\sigma}({\bf r}),\hat{H}(t)]|\Psi(t)}\\
        &=\int d\sigma\braket{\Psi(t)|[[\hat{\psi}_{\sigma}^{\dagger}({\bf r'})\hat{\psi}_{\sigma}({\bf r}),\hat{H}(t)],\hat{H}(t)]|\Psi(t)}+i\int d\sigma\braket{\Psi(t)|[\hat{\psi}_{\sigma}^{\dagger}({\bf r'})\hat{\psi}_{\sigma}({\bf r}),\dot{\hat{H}}(t)]|\Psi(t)}\\
        &\sim i\int d\sigma\braket{\Psi(t)|[\hat{\psi}_{\sigma}^{\dagger}({\bf r'})\hat{\psi}_{\sigma}({\bf r}),\dot{\hat{v}}_{\rm ext}(t)]|\Psi(t)},
    \end{split}
    \label{eq:twotderivative}
\end{equation}
where we used the notation of second quantization under the fermionic field operator $\hat{\psi}_{\sigma}({\bf r})$ and its adjoint $\hat{\psi}_{\sigma}^{\dagger}({\bf r})$. The first term of the second line in Eq.~\eqref{eq:twotderivative} vanishes due to the adiabatic assumption in Eq.~\eqref{eq:adia_eigen}. The final form is given by
\begin{equation}
    -\frac{\partial^2}{\partial t^2}\gamma({\bf r},{\bf r'},t)\sim \bigg[i\frac{\partial}{\partial t}(v_{\rm ext}({\bf r},t)-v_{\rm ext}({\bf r}',t))\bigg]\gamma({\bf r},{\bf r'},t).
\end{equation}
On the other hand, the adiabatic 1RDM is assumed to be $\partial_t^2\gamma_{\rm adia}({\bf r},{\bf r'},t)\sim 0$. Subtracting the second-order time derivative between $\gamma({\bf r},{\bf r'},t)$ and $\gamma_{\rm adia}({\bf r},{\bf r'},t)$ and substituting $\gamma({\bf r},{\bf r'},t)=\gamma_{\rm adia}({\bf r},{\bf r'},t)$ into the non-time-derivative term on the right-hand side of Eq. (B3), we obtain Eq.~\eqref{eq:deltagamma_2ndderiv}.
\twocolumngrid

\bibliographystyle{apsrev4-2}

\end{document}